# A Comprehensive Utility Function for Resource Allocation in Mobile Edge Computing

Zaiwar Ali[1], Sadia Khaf[2], Ziaul Haq Abbas[2], Ghulam Abbas[3], Lei Jiao[4], Amna Irshad[2], Kyung Sup Kwak[5] and Muhammad Bilal[6,*]

[1]Telecommunications and Networking (TeleCoN) Research Laboratory, GIK Institute of Engineering Sciences and Technology, Topi, 23640, Pakistan
[2]Faculty of Electrical Engineering, GIK Institute of Engineering Sciences and Technology, Topi, 23640, Pakistan
[3]Faculty of Computer Sciences and Engineering, GIK Institute of Engineering Sciences and Technology, Topi, 23640, Pakistan
[4]Department of Information and Communication Technology, University of Agder (UiA), Grimstad, 4898, Norway
[5]Department of Information and Communication Engineering, Inha University, Incheon, 22212, Korea
[6]Department of Computer and Electronics Systems Engineering, Hankuk University of Foreign Studies, Gyeonggi-do, 17035, Korea
*Corresponding Author: Muhammad Bilal. Email: m.bilal@ieee.org
Received: 19 August 2020; Accepted: 14 September 2020

**Abstract:** In mobile edge computing (MEC), one of the important challenges is how much resources of which mobile edge server (MES) should be allocated to which user equipment (UE). The existing resource allocation schemes only consider CPU as the requested resource and assume utility for MESs as either a random variable or dependent on the requested CPU only. This paper presents a novel comprehensive utility function for resource allocation in MEC. The utility function considers the heterogeneous nature of applications that a UE offloads to MES. The proposed utility function considers all important parameters, including CPU, RAM, hard disk space, required time, and distance, to calculate a more realistic utility value for MESs. Moreover, we improve upon some general algorithms, used for resource allocation in MEC and cloud computing, by considering our proposed utility function. We name the improved versions of these resource allocation schemes as comprehensive resource allocation schemes. The UE requests are modeled to represent the amount of resources requested by the UE as well as the time for which the UE has requested these resources. The utility function depends upon the UE requests and the distance between UEs and MES, and serves as a realistic means of comparison between different types of UE requests. Choosing (or selecting) an optimal MES with the optimal amount of resources to be allocated to each UE request is a challenging task. We show that MES resource allocation is sub-optimal if CPU is the only resource considered. By taking into account the other resources, i.e., RAM, disk space, request time, and distance in the utility function, we demonstrate improvement in the resource allocation algorithms in terms of service rate, utility, and MES energy consumption.

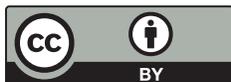

## 1 Introduction

With the exponential growth of the Internet of Things (IoT) and 5G technologies, highly advanced mobile applications, such as online games, face and speech recognition etc., are rapidly emerging [1]. Because of the computationally and communicationally expensive applications, the computation capability and battery life of a user equipment (UE) are normally insufficient [2]. Cloud computing can, to a certain extent, enhance the computing capacity of UEs, however, long distances between cloud and UEs lead to high latency and energy consumption [3,4]. Moreover, the demand for high definition and low latency mobile applications is increasing day by day. This tendency poses significant challenges to the existing mobile networks especially during the high traffic hours [5]. To address these problems, a new computing paradigm has emerged, known as mobile edge computing (MEC) [1,2]. MEC enables the cloud computing capabilities and storage services at the edge of a network, which enables UEs to execute their tasks more efficiently [3].

In MEC, two problems are of great importance, namely, task offloading and resource allocation [6]. This paper concerns the latter, which refers to the ability of a central control unit to assign UEs to mobile edge servers (MESs) and to allocate specific amount of resources to UE requests. Thus, the UEs execute their applications remotely in a more efficient way by using powerful MESs. The MESs, in return for allowing their resources to be used, obtain a reward in terms of a unit-less quantity, named as *utility* [7].

### 1.1 Related Work

According to Wu et al. [7], different individuals have different utility functions due to distinct needs towards a specific service. However, there are some general requirements in resource allocation problems in MEC, which must be considered in utility functions. Central processing unit (CPU), random access memory (RAM), hard disk, and the time for which these resources will be used are important parameters to be considered in a utility function. The authors in Fernandes et al. [8] consider CPU, RAM, and other hardware components to design a virtual machine (VM) scheduler, which schedules VMs in such a way to minimize the energy consumption of data centers. However, this approach is not a utility based approach and can be used only for cloud computing. The authors in [9,10] investigate the resource allocation problem and propose a deep learning approach to minimize the service time and efficiently allocate resources to UEs. However, the energy consumption and the utility function for MESs are not considered. The authors in Dlamini et al. [11] propose a computing-plus-communication energy model for resource allocation in MEC, where they use a hybrid-powered MES and switching techniques of transmission drivers to minimize MESs' energy consumption.

The authors in Li et al. [12] present a stochastic approach, using Lyapunov optimization technique, for wireless-powered MEC to minimize the energy consumption by optimizing the transmission power of MESs. The authors in Liu et al. [13] use the effect of dynamic energy variation to propose a dynamic game-based approach for resource allocation in wireless powered MEC. In their approach, the resources are optimally allocated by computing optimal transmission power and optimal task offloading. The authors in Kan et al. [14] consider the task offloading and resource allocation problems in MEC. The resource allocation problem is solved as a cost minimization problem. However, the work mainly focuses on task offloading, and for resource allocation the utility for MESs is ignored. Futhermore, the authors in [15,16] propose a VM placement algorithm for cloud computing. The main objective of their work is to minimize the energy consumption by minimizing the number of active servers. However, in a utility function realistic needs of UEs can be introduced to generalize the technique for MEC.

Most of the literature on resource allocation [17–20] in MEC and cloud computing considers only the CPU as resource requested by the UEs. However, hard disk space, RAM, and the time that an MES allocates to UEs for using resources are also important parameters but are ignored in dealing with UE requests.

Moreover, most of the related work considers the utility function as a uniform random variable [7] or only considers the requested CPU in utility function [21,22]. However, the utility function for MES must be proportional to the amount of all resources used by UEs.

*1.2 Novelty and Contribution*

In the literature, there are certain algorithms that employ the concept of utility function for resource allocation problem in MEC and cloud computing [1,22]. In this paper, by utilizing our comprehensive mathematical model for utility function and UE requests, we modify these algorithms and make them more comprehensive for MEC. To make the mathematical model more realistic and comprehensive, we consider CPU, RAM, disk space, required time, and the distance between MES and UE in the mathematical model for UE request and utility function. To the best of our knowledge, no such comprehensive mathematical model exists for UE request and utility function in MEC resource allocation. The novelty and contributions of our work are summarized as follows:

- We consider CPU, RAM, hard disk, and requested time that will be used by UEs, as a UE request in resource allocation problem in MEC. The existing schemes consider only CPU as UE request.
- A novel comprehensive utility function for MESs is developed which depends on all the above mentioned resources. It also depends on the distance between UE and MES, which ensures the dependency of utility on quality-of-service (QoS).
- By using our proposed mathematical model, we improve four well-known MES resource allocation algorithms, namely, Basic Over-provisioning (BO), Greedy Max (GM), Minimum Expand (MinExpand), and Power Minimum Expand (PowExpand).
- Our corresponding comprehensive algorithms, namely, Comprehensive Basic Overprovisioning (CBO), Comprehensive Greedy Max (CGM), Comprehensive Minimum Expand (CMinExpand), and Comprehensive Power Minimum Expand (CPowExpand), improve the utility for MESs and the service rate, and minimize the energy consumption of MESs.

The rest of the paper is structured as follow. Section II describes the mathematical model for the UE request and utility function. Section III presents the proposed comprehensive resource allocation schemes. Section IV presents simulation results and Section V concludes the paper.

## 2 System Model

When a UE offloads its task for execution to the central control unit of MEC, it is important to allocate resources of different MESs upon incoming UE's requests. In our system model, there are multiple MESs with different amounts of available resources, as shown in Fig. 1. It is a challenging task to select an MES and allocate its resource to incoming UE's requests to improve service rate, power efficiency, and utility simultaneously. In this paper, we propose utility based comprehensive algorithms for resources allocation. We consider the same multi-user multi-server scenario as used by Cardosa et al. [22]. The number of incoming requests is modeled according to the Poisson distribution. It is assumed that the system consists of a central control unit that detects the incoming UE requests, selects the optimal MES for each request, and allocates MES resources to these requests.

*2.1 The User Request*

We consider $n$ servers and $m$ UE requests. The request matrix is modelled in two parts: (i) The requested resources matrix **Q**, and (ii) The distance matrix **D**. The resource matrix **Q** is given by:

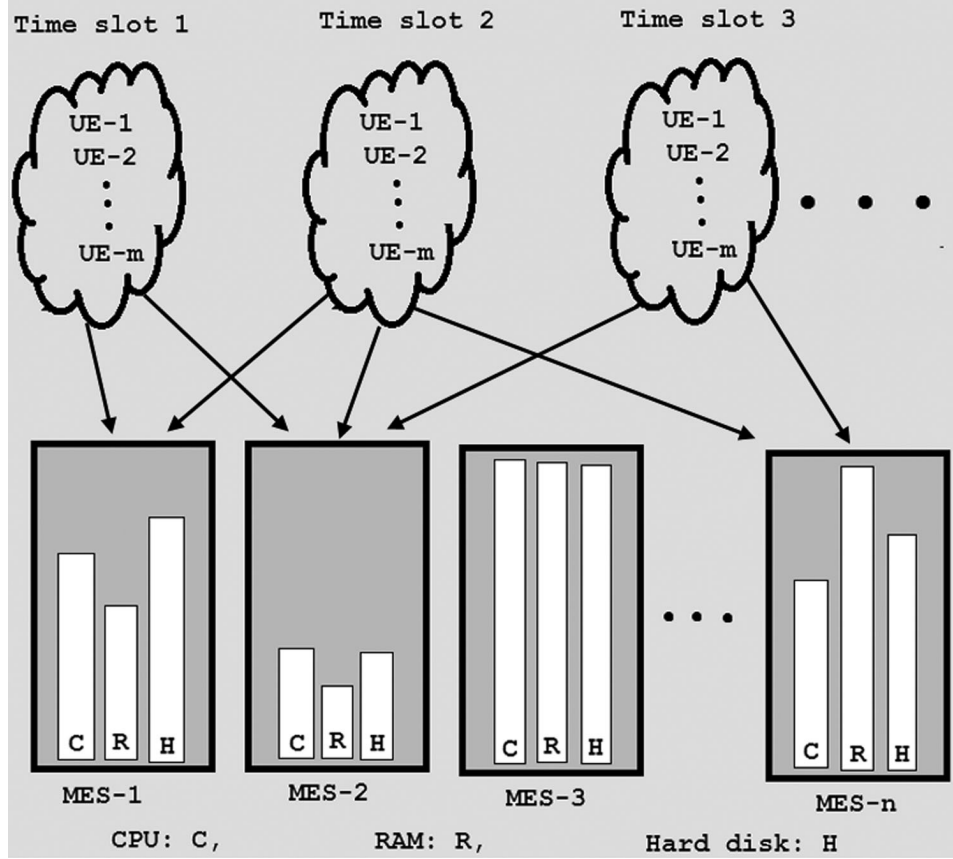

**Figure 1:** MESs with different amounts of available resources

$$Q = \left[ c_{min}^T, \ c_{\max}^T, \ r_{min}^T, \ r_{\max}^T, \ h^T, \ t^T \right], \tag{1}$$

where $c_{min}^T$ and $c_{max}^T$, respectively, represent the vectors of the minimum and the maximum amount of CPU in $m$ UE requests. $r_{min}^T$ and $r_{max}^T$, respectively, represent the vectors of the minimum and the maximum amount of RAM in UE requests. $h^T$ is the vector of the amount of disk space in UE requests, and $t^T$ is the vector of the number of time slots specified in the UE requests for which these resources are needed. We assume that this required time for using the resources of an MES is enough for execution of UE's task. Alternatively, the required time can be exceeded in the next UE request. For example, if the requested time is not enough for the execution of the given task then UE sends another request for the remaining task. The disk space is not modeled as a variable resource because, in most real-word scenarios, an application needs a fixed amount of disk space for its operation. Thus, there are six important parameters that a UE sends to the central control unit as its request. Therefore, the order of the matrix $\mathbf{Q}$ is $m \times 6$, where $m$ is the total number of incoming UE requests. The vectors in $\mathbf{Q}$ can be written as:

$$c_{min} = [c_{1_{min}}, c_{2_{min}}, c_{3_{min}}, \ldots, c_{m_{min}}], \tag{2}$$

$$c_{max} = [c_{1_{max}}, c_{2_{max}}, c_{3_{max}}, \ldots, c_{m_{max}}], \tag{3}$$

$$r_{min} = [r_{1_{min}}, r_{2_{min}}, r_{3_{min}}, \ldots, r_{m_{min}}], \tag{4}$$

$$\boldsymbol{r}_{max} = [r_{1_{max}}, r_{2_{max}}, r_{3_{max}}, \ldots, r_{m_{max}}], \tag{5}$$

$$\boldsymbol{h} = [h_1, h_2, h_3, \ldots, h_m], \tag{6}$$

$$\boldsymbol{t} = [t_1, t_2, t_3, \ldots, t_m], \tag{7}$$

where $c_{1_{min}}, c_{1_{max}}, r_{1_{min}}, r_{1_{max}}, h_1$, and $t_1$ represent the minimum CPU, the maximum CPU, the minimum RAM, the maximum RAM, the disk space, and the time requested by UE 1, respectively. In general for UE $j$, we can denote these requested resources as $c_{j_{min}}, c_{j_{max}}, r_{j_{min}}, r_{j_{max}}, h_j$, and $t_j$. The dimensions of vectors $\boldsymbol{c}_{min}, \boldsymbol{c}_{max}, \boldsymbol{r}_{min}, \boldsymbol{r}_{max}, \boldsymbol{h}$, and $\boldsymbol{t}$ are $1 \times m$.

The distance matrix $\boldsymbol{D}$, containing the distances of $m$ users from $n$ MESs, is given by:

$$\boldsymbol{D} = \begin{bmatrix} d_{11} & d_{12} & \cdots & d_{1n} \\ d_{21} & d_{22} & \cdots & d_{2n} \\ \vdots & \vdots & \ddots & \vdots \\ d_{m1} & d_{m2} & \cdots & d_{mn} \end{bmatrix}, \tag{8}$$

where $d_{11}$ is the distance between UE 1 and MES 1. Similarly, $d_{mn}$ is the distance between UE $m$ and MES $n$. In general, $d_{jk}$ represents the distance of UE $j$ from MES $k$.

### 2.2 The Proposed Comprehensive Utility Function

*Utility* is the reward that an MES receives for allowing its resources to be used by UEs. Since MESs provide their CPU, RAM, and hard disk for a specific period of time to UEs, the utility in this paper depends on the amount of CPU, RAM, and disk space requested by the UE, the time for which a UE has requested these resources, and the distance between the UE and the MES. Therefore, the utility function is directly proportional to the requested resources and time. The QoS that the UE receives is inversely proportional to the distance between the UE and the MES because of the network conditions, the transmission range of the UE and the MES, the frequent disconnections, and the latency in communication. The utility is, therefore, given as:

$$u_{jk} = \frac{(\gamma_1 c_j + \gamma_2 r_j + \gamma_3 h_j)\gamma_4 t_j}{d_{jk}}. \tag{9}$$

Here, $u_{jk}$ is the utility for MES $k$ for serving UE $j$. $c_j$, $r_j$, and $h_j$, respectively, denote the amount of CPU, RAM and hard disk space allocated to UE $j$ by MES $k$, $t_j$ is the time for which these resources are allocated, and $d_{jk}$ is the distance between UE $j$ and MES $k$. The requested resources are weighted and normalized by the unit balancing and weighting coefficients ($\gamma_1$, $\gamma_2$, and $\gamma_4$, respectively) as:

$$\gamma_1 = \frac{w_1}{c_{total}}, \tag{10}$$

$$\gamma_2 = \frac{w_2}{r_{total}}, \tag{11}$$

$$\gamma_3 = \frac{w_3}{h_{total}}, \tag{12}$$

where $c_{total}$, $r_{total}$, and $h_{total}$ are the combined total CPU, RAM, and hard disk space of all the servers, respectively, and $w_1$, $w_2$, and $w_3$ are the weighting coefficients. For example, the value of $w_1$ shows the relative contribution of CPU to the utility function. These coefficients can be adjusted to represent the expensiveness of different resources. Similarly, to keep the utility function unitless, $t_j$ and $d_{jk}$ are normalized with respect to the unit balancing and weighting coefficient $\gamma_4$, i.e.,

$$\gamma_4 = \frac{d_{max}}{t_{max}}, \tag{13}$$

where $d_{max}$ is the maximum distance within which all MESs and UEs can operate, and $t_{max}$ is the maximum time that a UE is allowed to request. The upper threshold on time is configured such that a UE cannot occupy all the resources and prevent other UEs from obtaining the service.

The utility for each MES is different from other MESs for serving the UE requests because of the UE's location. An MES receives higher utility by serving the nearby UE than by serving a far UE because a user closer to the MES has better QoS and pays a higher premium for it. The utility of $n$ MESs for entertaining the same UE $j$ can, thus, be represented by a vector $\mathbf{u}$ as:

$$\mathbf{u} = [u_{j1}, u_{j2}, u_{j3}, \ldots, u_{jn}]. \tag{14}$$

Similarly, the utility matrix $\mathbf{U}$ for $m$ UEs and $n$ MESs can be written as:

$$\mathbf{U} = \begin{bmatrix} u_{11} & u_{12} & \cdots & u_{1n} \\ u_{21} & u_{22} & \cdots & u_{2n} \\ \vdots & \vdots & \ddots & \vdots \\ u_{m1} & u_{m2} & \cdots & u_{mn} \end{bmatrix}, \tag{15}$$

where $u_{11}$ is the utility of UE 1 at MES 1, $u_{mn}$ is the utility of UE $m$ at MES $n$.

According to a UE request, the minimum and the maximum utilities that an MES can achieve by serving the request are represented as:

$$u_{jk_{min}} = \frac{(\gamma_1 c_{j_{min}} + \gamma_2 r_{j_{min}} + \gamma_3 h_j)\gamma_4 t_j}{d_{jk}}, \tag{16}$$

and

$$u_{jk_{max}} = \frac{(\gamma_1 c_{j_{max}} + \gamma_2 r_{j_{max}} + \gamma_3 h_j)\gamma_4 t_j}{d_{jk}}, \tag{17}$$

where $u_{jk_{min}}$ and $u_{jk_{max}}$, respectively, represent the minimum and maximum utilities for MES $k$ from UE $j$. $c_{j_{min}}$ and $c_{j_{max}}$, respectively, represent the minimum and maximum of the CPU resources requested by UE $j$. Moreover, $r_{j_{min}}$ and $r_{j_{max}}$, respectively, represent the minimum and maximum amounts of RAM requested by UE $j$.

From a set of $n$ available MESs, we find an MES $k$, such that $k \in \{1, 2, 3, \ldots, n\}$, for which the utility of a UE $j$ is maximum. This MES $k$ is called the optimal MES, $s_j^*$, for UE $j$, and is represented as:

$$s_j^* = \max_{k \in \{1,2,\ldots,n\}} u_{jk}, \qquad j = 1, 2, \ldots, m. \tag{18}$$

### 2.3 Feasibility of Servers

An MES $k$ is feasible for UE $j$ if the currently available resources of MES $k$ exceed the resources requested by UE $j$. A feasibility vector $\mathbf{f}_j$ for UE $j$ is given by:

$$\boldsymbol{f_j} = [f_{j1}, f_{j2}, f_{j3}, \ldots, f_{jn}], \tag{19}$$

where $f_{j1}$ is the feasibility status of MES 1 for UE $j$, and so on. Formally, $f_{j1} = 1$ means that MES 1 is feasible for UE $j$ and $f_{j1} = 0$ means that MES 1 is not feasible for UE $j$. The dimension of the feasibility vector $\boldsymbol{f_j}$ is $1 \times n$. The feasibility $f_{jk}$ is calculated as:

$$f_{jk} = \begin{cases} 1 & c_{j_{\min}} \leq c_{k_{av}} \text{ and } r_{j_{\min}} \leq r_{k_{av}} \text{ and } h_j \leq h_{k_{av}} \\ 0 & \text{otherwise} \end{cases}, \tag{20}$$

where $c_{j_{\min}}$, $r_{j_{\min}}$, and $h_j$ are the minimum resources requested by UE $j$. $c_{k_{av}}$, $r_{k_{av}}$, and $h_{k_{av}}$ are the resources currently available at MES $k$. This feasibility vector represents the feasibility of servers for each UE. For example, $\boldsymbol{f_j} = [1, 1, 0, 0, 0, 1, 1, 1, 0, 0]$ means that MES 1, 2, 6, 7, and 8 are feasible for UE $j$. We can only serve UE $j$ at one of the feasible servers.

### 2.4 Energy Aware MES Priority

The energy consumed by each MES for keeping itself ON (active) is different from other MESs because of its capacity and type of hardware used. The energy consumed per unit time in keeping the MES $k$ ON is represented by $E_k$. The total resources of MES $k$ define its capacity $C\_P_k$ as:

$$C\_P_k = \gamma_1 c_{k_{total}} + \gamma_2 r_{k_{total}} + \gamma_3 h_{k_{total}}, \tag{21}$$

where $c_{k_{total}}$, $r_{k_{total}}$, and $h_{k_{total}}$ represent the total CPU resources, RAM, and hard disk space of MES $k$, respectively. The energy consumption per unit capacity, $p_k$, of MES $k$ can be written as:

$$p_k = \frac{E_k}{C\_P_k}. \tag{22}$$

The MESs are sorted in increasing order of $p_k$. An MES with lower value of $p_k$ shows that it has a higher capacity for serving larger UE requests while its energy consumption for keeping it ON is low. Hence, it should be employed more often than other MESs with higher value of $p_k$.

An MES that is already in the ON state should be prioritized for entertaining new UE requests, whereas the MESs that are in the idle state should be avoided to activate as long as possible to conserve energy. Our proposed algorithms use the more profitable and already active servers first, and only activate the less profitable servers if the currently active servers are not enough to handle the incoming traffic. Therefore, we introduce the penalty in the utility function as:

$$u'_{jk} = \begin{cases} u_{jk} & s_k = 1 \\ u_{jk} - \gamma_5 p_k & s_k = 0 \end{cases}, \tag{23}$$

where $s_k = 1$ indicates that MES $k$ is ON, and $s_k = 0$ implies that MES $k$ is idle. The unit balancing coefficient $\gamma_5$ is given as:

$$\gamma_5 = \frac{w_5}{e_{max}}, \tag{24}$$

where $w_5$ is the weighting coefficient and its value can be adjusted to make the threshold for activating an idle server higher or lower, and $e_{max}$ is the sum of the $E_k$ for all servers.

The energy consumption due to the usage of CPU, RAM, and disk space depends on the instruction type and architecture of the system used in an MES. We assume linear relation between the energy consumption and usage of CPU, RAM, and disk space [23]. The energy consumption due to CPU usage, RAM usage, and disk space usage of MES $k$, i.e., $E_{ck}$, $E_{rk}$, and $E_{hk}$, respectively, can be given as:

$$E_{ck} = E_{ckmin} + (E_{ckmax} - E_{ckmin})G_c, \tag{25}$$

$$E_{rk} = E_{rkmin} + (E_{rkmax} - E_{rkmin})G_r, \tag{26}$$

$$E_{hk} = E_{hkmin} + (E_{hkmax} - E_{hkmin})G_h, \tag{27}$$

where $E_{ckmin}$, $E_{rkmin}$, and $E_{hkmin}$, respectively, represent the energy consumption when the CPU, RAM, and disk space are not in use. Similarly, $E_{ckmax}$, $E_{rkmax}$, and $E_{rkmax}$ represent the energy consumption when the CPU, RAM, and disk space, respectively, are fully utilized. $G_c$, $G_r$, and $G_h$ are the utilization of CPU, RAM, and disk space, respectively. The total energy consumption of MES $k$, $E_{k_{total}}$, is therefore calculated as:

$$E_{k_{total}} = E_k \cdot t_{k_{active}} + E_{ck} + E_{rk} + E_{hk}, \tag{28}$$

where $t_{k_{active}}$ is the total time for which the MES $k$ was active. The total energy consumption of all the MESs, $E_{total}$, can be written as:

$$E_{total} = \sum_{k=1}^{n} E_{k_{total}}. \tag{29}$$

## 3 Resource Allocation Schemes

In MEC, the central control unit assigns MESs and the resources of MESs to different UE's requests. To select the best MES for a specific UE and to allocate the resources according to the request of UE are challenging. In the literature, there are certain general algorithms used for resource allocation in cloud computing and MEC which only consider the CPU as UE requests or a random utility function independent of UE requests. We improve the well-known MES resource allocation algorithms [22] by considering our comprehensive mathematical formulation for UE request and utility function of MES. In this section, we explain our proposed comprehensive algorithms.

### 3.1 Comprehensive Basic Over-Provisioning

The idea behind basic over-provisioning scheme (BO) used in Cardosa et al. [22] is about serving the UEs on a first-come first-served basis by allocating the maximum requested CPU to them. The BO scheme keeps fitting VMs into the first available MES until the server is left with 10% of its maximum CPU. Following the same approach, our proposed comprehensive BO (CBO) scheme determines the available capacity at the servers one by one in the decreasing order of their profitability and creates the VM of the incoming UE requests at the first available MES.

Algorithm 1 explains CBO, which is different from BO in the sense that it also considers RAM and disk space instead of considering CPU only. In addition, it also considers the time for which a UE has requested these resources and creates the VM for that particular time. When the time that the UE requested has elapsed, the VM is deleted automatically and the resource usage of the MES is adjusted accordingly. CBO is similar to BO in the sense that it does not differentiate between UE requests in terms of their utilities.

**Algorithm 1:** Comprehensive Basic Over-provisioning

| | |
|---|---|
| 1: | Sort all MESs into increasing order of $p_k$ |
| 2: | **for** time $i = 1, 2, \ldots, t$ **do** |
| 3: |    **for** UE $j = 1, 2, \ldots, m$ **do** |
| 4: |       **for** MES $k = 1, 2, \ldots, n$ **do** |

**Algorithm 1 (continued).**

| | |
|---|---|
| 5: | **if** $c_{j_{\max}} < 0.9c_{k_{av}}$ And $r_{j_{\max}} < 0.9r_{k_{av}}$ And $h_j < 0.9h_{k_{av}}$ **then** |
| 6: | Activate MES $k$ if it is idle |
| 7: | Create VM for UE $j$ at MES $k$ for time $t_j$ |
| 8: | Allocate maximum requested resources |
| 9: | Update resource usage and overall utility using Eq. (1) and Eq. (15) |
| 10: | *break;* |
| 11: | **end if** |
| 12: | **end for** |
| 13: | **end for** |
| 14: | Check distance of all existing UEs from their respective MESs |
| 15: | Disconnect UEs that go out of the range of MES |
| 16: | Update resource usage and overall utility |
| 17: | Bring the MESs with zero usage to power save mode |
| 18: | **end for** |

### 3.2 Comprehensive Greedy Max

Serving the UE requests regardless of the utility that they offer is disadvantageous for MESs. The Greedy Max (GM) algorithm [22] tackles this problem. However, GM considers only CPU as UE request and, therefore, the utility function only depends on the requested CPU. We propose a Comprehensive Greedy Max (CGM) algorithm in which we consider CPU, RAM, hard disk, required time, and distance as UE request. Therefore, we model our comprehensive utility function depending on these realistic values in a UE request. Contrary to CBO, CGM first sorts the incoming UE requests in the decreasing order of their utility. For instance, if there are three UE requests in the first time slot, CBO scheme allocates the maximum requested resources to each of them regardless of any priority, whereas CGM sorts them into the decreasing order of their maximum requested resources (maximum utility), and then allocates the maximum requested resources to them. CGM is presented in Algorithm 2.

**Algorithm 2:** Comprehensive Greedy Max

| | |
|---|---|
| 1: | Sort all MESs into increasing order of $p_k$ |
| 2: | **for** time $i = 1, 2, \ldots, t$ **do** |
| 3: | Sort all incoming UEs into decreasing order of requested resources |
| 4: | **for** UE $j = 1, 2, \ldots, m$ **do** |
| 5: | **for** MES $k = 1, 2, \ldots, n$ **do** |
| 6: | **if** $c_{j_{\max}} < 0.9c_{k_{av}}$ And $r_{j_{\max}} < 0.9r_{k_{av}}$ And $h_j < 0.9h_{k_{av}}$ **then** |
| 7: | Activate MES $k$ if it is idle |

(*Continued*)

**Algorithm 2 (continued).**

| | |
|---|---|
| 8: | Create VM for UE *j* at MES *k* for time $t_j$ |
| 9: | Allocate maximum requested resources |
| 10: | Update resource usage and overall utility using Eq. (1) and Eq. (15) |
| 11: | *break;* |
| 12: | **end if** |
| 13: | **end for** |
| 14: | **end for** |
| 15: | Check distance of all existing UEs from their respective MESs |
| 16: | Disconnect UEs that go out of the range of MES |
| 17: | Update resource usage and overall utility |
| 18: | Bring the MESs with zero usage to power save mode |
| 19: | **end for** |

In the case of high traffic, the difference between the performance of CBO and CGM is clear when the MESs start filling up and some of the UEs are denied services. The UEs that are denied services, in case of CGM scheme, will always be the ones that offered the lowest utility.

### 3.3 Comprehensive Minimum Expand

The problem with CBO and CGM schemes is that in peak hours, they keep allocating the maximum requested resources to certain UEs and keep denying service to all others. Comprehensive Minimum Expand (*CMinExpand*) solves this problem by allocating the minimum requested CPU, RAM, and hard disk space to UEs for the complete time that they requested the resources for. Later, the scheme allocates them more resources only if there is still room available at the MES after giving the minimum resources to all incoming UE requests. Thus, the scheme allocates minimum resources to all the UEs and then expands to maximum requested resources. Therefore, this scheme is called *CMinExpand*. In this way, *CMinExpand* gives service to a lot more UEs in peak hours than CBO and CGM. *CMinExpand* is the extended version of *MinExpand* algorithm [22], which considers only CPU as UE request.

*CMinExpand* is also greedy in nature because it follows the same principal as the CGM scheme for sorting when it is expanding the existing VMs. It allocates the minimum resources to all UEs, but then expands them due to their decreasing utility. In this manner, the more profitable VMs get expanded first. The expansion takes place until there is room on the MES. Thus, if some of the VMs do not get expanded because of the server running out of resources to allocate, they will always be the least profitable VMs. The expansion takes place until the UEs' requested maximum resources have reached or the server runs out of resources, whichever happens first. *CMinExpand* is presented in Algorithm 3.

**Algorithm 3:** Comprehensive Minimum Expand

| | |
|---|---|
| 1: | Sort all MESs into increasing order of $p_k$ |
| 2: | **for** time $i = 1, 2, \ldots, t$ **do** |
| 3: |   **for** UE $j = 1, 2, \ldots, m$ **do** |
| 4: |     **for** MES $k = 1, 2, \ldots, n$ **do** |
| 5: |       **if** $c_{j_{\max}} < 0.9 c_{k_{av}}$ And $r_{j_{\max}} < 0.9 r_{k_{av}}$ And $h_j < 0.9 h_{k_{av}}$ **then** |
| 6: |         Activate MES $k$ if it is idle |
| 7: |         Create VM for UE $j$ at MES $k$ for time $t_j$ |
| 8: |         Allocate maximum requested resources |
| 9: |         Update resource usage and overall utility using Eq. (1) and Eq. (15) |
| 10: |         *break;* |
| 11: |       **end if** |
| 12: |     **end for** |
| 13: |   **end for** |
| 14: |   **for** MES $k = 1, 2, \ldots, n$ **do** |
| 15: |     Sort all VMs at MES $k$ into decreasing order of their $u_{j_{\max}}$ |
| 16: |     **while** $c_{k_{av}} > 0.1 c_k$ **do** |
| 17: |       **for** sorted VMs $j = 1, 2, \ldots$ at MES $k$ **do** |
| 18: |         Expand VM $j$ to its maximum |
| 19: |         Update overall utility and resource usage |
| 20: |       **end for** |
| 21: |     **end while** |
| 22: |   **end for** |
| 23: |   Check distance of all existing UEs from their respective MESs |
| 24: |   Disconnect UEs that go out of the range of MES |
| 25: |   Update resource usage and overall utility |
| 26: |   Bring the MESs with zero usage to power save mode |
| 27: | **end for** |

### 3.4 Comprehensive Power Minimum Expand

The disadvantage of all the above described allocation schemes is that they prioritize service provisioning to UEs regardless of any energy consumption constraints on the MES side. If the most profitable server becomes full, they turn the next server ON without setting any utility threshold for it. Comprehensive Power Minimum Expand (*CPowExpand*) attempts to find a balance between providing service to UEs and the energy consumption of the MES. *CPowExpand* is the improved version of *PowExpand* algorithm [22].

*CPowExpand* sets a certain threshold for bringing a server to ON state from idle state and employs the penalized utility function given by Eq. (23). If a server is already in the ON state, the utility is the same as before. However, if a server is in the idle state for energy saving, it subtracts a penalty term from the utility function so that the servers that are already in the ON state are prioritized over the idle server as long as there is resource available on them. As before, the difference between the *PowExpand* and our proposed *CPowExpand* is that the latter considers RAM, disk space, time, distance from the MES as part of the UE request, and the utility function for MES. *CPowExpand* is presented in Algorithm 4.

**Algorithm 4:** Comprehensive Power Minimum Expand

1: Sort all MESs into increasing order of $p_k$
2: **for** time $i = 1, 2, \ldots, t$ **do**
3:   **for** UE $j = 1, 2, \ldots, m$ **do**
4:     **for** MES $k = 1, 2, \ldots, n$ **do**
5:       **if** $c_{j_{max}} < 0.9 c_{k_{av}}$ And $r_{j_{max}} < 0.9 r_{k_{av}}$ And $h_j < 0.9 h_{k_{av}}$ **then**
6:         **if** MES $k$ is idle **then**
7:           Compute $u'_{j_{min}}$ using Eq. (23)
8:           **if** $u'_{j_{min}} > 0$ **then**
9:             Activate MES $k$
10:           **else**
11:             break;
12:           **end if**
13:         **end if**
14:         Create VM for UE $j$ at MES $k$ for time $t_j$
15:         Allocate maximum requested resources
16:         Update resource usage and overall utility using Eq. (1) and Eq. (15)
17:         break;
18:       **end if**
19:     **end for**
20:   **end for**
21:   **for** MES $k = 1, 2, \ldots, n$ **do**
22:     Sort all VMs at MES $k$ into decreasing order of their $u_{j_{max}}$
23:     **while** $c_{k_{av}} > 0.1 c_k$ **do**
24:       **for** sorted VMs $j = 1, 2, \ldots$ at MES $k$ **do**

| | Algorithm 4 (continued). |
|---|---|
| 25: |       Expand VM *j* to its maximum |
| 26: |       Update overall utility and resource usage |
| 27: |     **end for** |
| 28: |   **end while** |
| 29: | **end for** |
| 30: | Check distance of all existing UEs from their respective MESs |
| 31: | Disconnect UEs that go out of the range of MES |
| 32: | Update resource usage and overall utility |
| 33: | Bring the MESs with zero usage to power save mode |
| 34: | **end for** |

The disadvantage of CPowExpand is that some UEs may be denied services if the utility they offer is too small as compared to the threshold of activating an MES for them. The parameter $\gamma_5$ can be adjusted for a stricter or lighter emphasis on threshold.

The implementation complexity of our improved schemes is the same as for their corresponding existing schemes because every scheme has to find the best allocation option for *m* UEs with *n* available MESs. There are *nm* total possible options from which an allocation scheme selects the best option in terms of energy consumption, service rate, and utility for MES. From simulation results, we can observe that our schemes select the best allocation option by consuming low energy and getting high service rate and utility.

## 4 Performance Evaluation

We use MATLAB (R2019a) to evaluate the performance of the proposed comprehensive algorithms in comparison with the benchmark algorithms [22], namely, BO, GM, *MinExpand*, and *PowExpand*. The request arrival is modeled as a Poison process with mean 5. The results of the resource allocation schemes are recorded for 1000 time slots. The amount of CPU, RAM, and disk space at MES are taken as Normal distribution with mean and variance 15 and 5; 10 and 2; and 25 and 5, respectively. The energy consumption per unit time to keep an MES ON is taken proportional to the amount of resources of an MES. The coverage range of an MES is assumed to be up to 800 m. The distance between users and MES is taken as a uniform random variable with $d \in [1, 1000]$. The unit balancing coefficients $\gamma_1$, $\gamma_2$, $\gamma_3$, and $\gamma_4$ are considered as 0.4, 0.25, 0.25, and 0.1, respectively, since we assume that the expensiveness of CPU is greater than those of RAM and disk space. We also assume that the maximum distance between UE and MES is 1 km and the maximum time slots a UE can request is 10 units. The utility, service rate, and energy consumption of MESs are chosen as performance metrics. The effects of varying the UE requests, and the total number of MESs are observed.

### 4.1 Effect of Varying Traffic

Fig. 2a presents a comparison of the utility for all the algorithms. The solid lines show the utility of the proposed comprehensive algorithms, whereas the dashed lines show the utility of the algorithms that only consider CPU. It is clear that the utility of the comprehensive algorithms is always higher than the benchmark algorithms because the benchmark algorithms allocate resources proportional to CPU only and face disconnections when the MES does not have enough other resources to serve the UE's request. The comprehensive algorithms use the utility function to evaluate incoming UE's requests and prioritize the more profitable ones, hence, generating a higher utility.

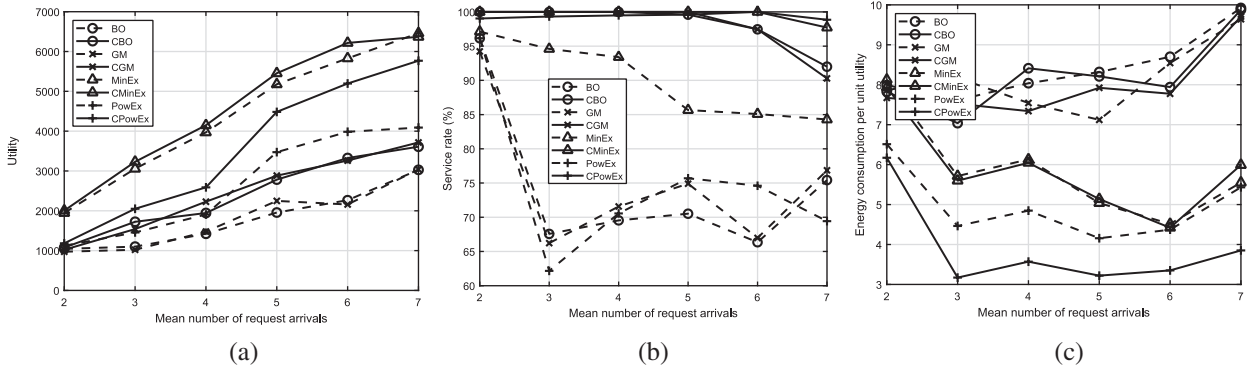

**Figure 2:** a) Utility under varying traffic scenarios, b) Service rate under varying traffic scenarios, c) Energy consumption per unit utility under varying traffic scenarios

Fig. 2b shows the comparison of the service rate for all the algorithms in low and high traffic. The solid lines represent the service rate of the comprehensive algorithms whereas the dashed lines show the service rate of the benchmark algorithms that only consider CPU. When very low traffic is injected, the service rates of all algorithms range from 90% to 99% because the available resources are enough to provide services to the incoming UEs. The service rates of all algorithms start decreasing when the traffic density starts increasing because the available resources are insufficient. Under very high traffic conditions, the service rate of the proposed comprehensive algorithms still remains above 90%. The benchmark algorithms face significant degradation in service rate even under medium traffic. This shows the importance of considering the heterogeneous nature of applications and the requested resources at deeper level.

Fig. 2c shows the energy consumption per unit utility for all the algorithms against different levels of incoming traffic. All the algorithms perform reasonably in low traffic conditions and their energy consumption per unit utility is quite low. However, under high traffic conditions, the algorithms that allocate the maximum requested resources to the users have a very high energy consumption per unit utility ratio. This represents that the energy consumption of non-comprehensive algorithms is very high for a relatively low level of utility, which means that for the same utility, our comprehensive algorithms will consume less energy.

### 4.2 Effect of Varying Total Number of Servers

In the previous subsection, we evaluated the impact of varying traffic on utility, service rate, and energy consumption of MES. In this subsection, we discuss the impact of varying the total number of servers available for the UE requests. Figs. 3a and 3b show the utility and service rate, respectively, of all the algorithms under varying number of total MESs available. When the number of servers is insufficient to serve all requests, both utility and service rate of all the algorithms is comparable. The difference in performance becomes apparent once there are enough MESs for all UE requests. For 10 MESs, we get high service rate with high utility under traffic with mean = 5 for all the algorithms. However, for more than 10 MESs, the comprehensive algorithms yield 100% service rate and high utility, whereas BO, GM, *MinExpand*, and *PowExpand* work at sub-optimal levels and achieve 60% to 90% service rate at best.

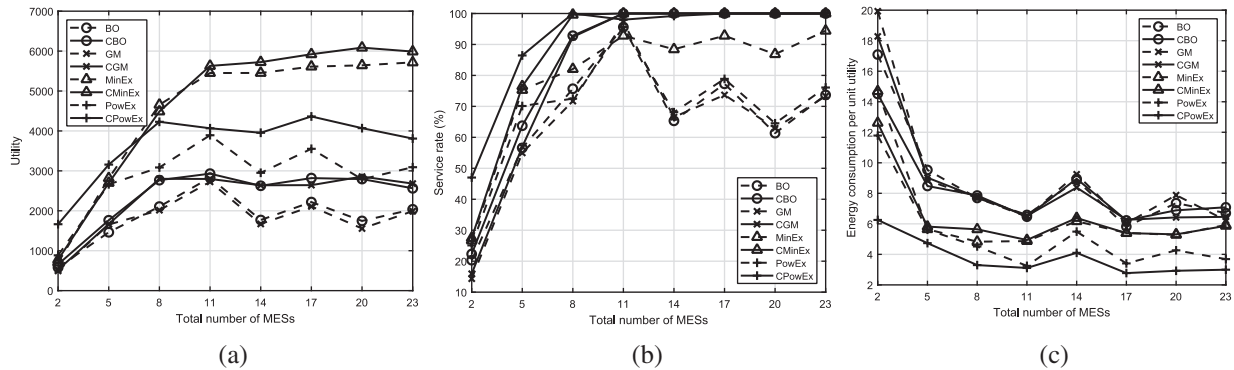

**Figure 3:** a) Utility under varying number of MESs, b) Service rate under varying number of MESs, c) Energy consumption per unit utility for different number of MESs

Fig. 3c shows the energy consumption per unit utility as a function of the number of servers. When the number of servers increases, the utility as well as the energy consumption of all the algorithms increases. Energy consumption per unit utility provides a meaningful measure of performance to compare different algorithms. It is clear that our comprehensive algorithms outperform the benchmark algorithms because of the higher utility of comprehensive algorithms for the same amount of energy consumption.

### *4.3 Discussions*

The simulation results show reasonable performance improvement of all the proposed comprehensive algorithms under varying traffic scenarios and varying number of MESs. The results highlight the importance of comprehensive utility function, especially when there are enough servers or low traffic, and all UE requests can be serviced. The proposed comprehensive algorithms are, in general, enough to be applied in most scenarios and the weights assigned to different resources can be adjusted according to the utility plan and applications. The performance improvements are achieved because the requested resources are considered at deeper level in our comprehensive approach. However, at present, we have not considered UE mobility and handover of UE requests from one MES to another. Therefore, as a future work, we will consider the VM migration for the mobile UEs.

## 5 Conclusion

For resource allocation in MEC, most literature considers CPU as the only resource that UEs can request. However, UEs can be denied services if servers run out of RAM or disk space while still having sufficient CPU resources to serve the UE's requests. In this paper, we have presented a comprehensive utility function that considers all realistic resources needed by UEs, including CPU, RAM, storage, required time, and the distance between MES and UE. We have improved the service rate and utility of several existing MES resource allocation schemes by incorporating our comprehensive utility function. We have also minimized the energy consumption of MESs, proposed a UE request structure and modeled the utility function on the information provided by UEs. The proposed utility function is used mainly for two purposes, i.e., choosing the optimal MES, and assigning different priorities to UE requests. We achieve significant improvement in service rate and utility while maintaining a low energy consumption for MESs. The results suggest that considering the heterogeneous nature of applications and the resources at deeper level can lead to high utility and service rate for resource allocation in MEC. As a future work, we aim to consider UE mobility to further optimize resource allocation in MEC.

**Funding Statement:** This work was supported by National Research Foundation of Korea-Grant funded by the Korean Government (Ministry of Science and ICT) -NRF-2020R1AB5B02002478.

**Conflicts of Interest:** The authors declare that they have no conflicts of interest to report regarding the present study.